\documentclass[seceq]{ptptex}
\usepackage{type1cm, 
hyperref}

\newcommand{\be}{\begin{equation}}
\newcommand{\ba}{\begin{eqnarray}}
\newcommand{\ea}{\end{eqnarray}}
\newcommand{\ee}{\end{equation}}

\newcommand{\Ncal}{\cal{N}}
\newcommand{\Ocal}{\cal{O}}
\DeclareMathOperator*{\Tr}{{\rm Tr}}

\title{
Classical c=1 Tachyon Scattering and 1/2 BPS Correlators}

\author{
Ta-Sheng {\sc Tai}\footnote{E-mail: tasheng@hep-th.phys.s.u-tokyo.ac.jp}}
\inst{
Department of 
Physics, University of Tokyo, Tokyo 113-0033, Japan}

\abst{
We study the correlator of chiral primary operators in 
$\Ncal$=4 super Yang-Mills theory in large $N$ limit. 
Through the free fermion picture, we map 
the gauge group rank and R-charges in SYM to 
the 
Fermi level and tachyon momenta, respectively, in the c=1 matrix model. 
By doing so, it is seen that half-BPS correlators are 
reproduced by tree-level tachyon 
scattering amplitudes.}

\newpage

\begin{document}

\maketitle

\section{Introduction and summary}

As is pointed out by Okuyama in Ref. 1), it is possible to use 
the Das-Jevicki-Sakita term in computing the 
two-point function of chiral primary operators in $\Ncal$=4 SYM. 
This elucidates the relation with the two-dimensional Yang-Mills theory.
Motivated by Ref. 4), in which 2d YM is related to the c=1 matrix model 
through the collective field theory established by Das and Jevicki, 
it is thus tempting to formulate a certain correspondence between the 
c=1 matrix model and $\Ncal$=4 SYM.

We find that the ``S-matrix'' 
structure in extremal 
two-point functions 
is essentially the same as that of the c=1 matrix model 
up to a phase and non-perturbative terms; 
that is, they are diagonalized by the relativistic fermion basis. 
In large $N$ limit, 
these half-BPS correlators 
can then be reproduced by the tree-level c=1 tachyon scattering, where 
non-perturbative effect is dropped out. 
This is carried out by mapping 
the gauge group rank $N$ and R-charges in SYM to 
the Fermi level $\mu$ and tachyon momenta, respectively.

The reason can be intuitively 
understood as follows. 
Since half-BPS chiral primary operators correspond to certain $N$-fermion 
quantum mechanical states in a harmonic oscillator potential, 
we are thus effectively comparing two kinds of ground Fermi liquids 
in the phase space, i.e. on the SYM side, it is a disk of radius 
$\sqrt{2N}$, while in the c=1 case, the profile is determined by 
the hyperbola $p^2-x^2\le -2\mu$. This makes clear $\mu\leftrightarrow N$. 
Also, by means of the bosonization, 
the map between R-charges and tachyon momenta can be accounted for 
due to their relation to 
momenta of the aforementioned 2d relativistic fermions.

The outline of this note is as follows. 
In $\S$2, we recall some basics of the complex matrix model. 
In $\S$3, we briefly review the c=1 matrix model and 
identify classical tachyon scattering amplitudes with half-BPS correlators. 
\\
\\

\section{Complex matrix model and 1/2-BPS correlators}

We first recall some ingredients in the computation of 
extremal correlators in $\Ncal$=4 SYM, following Ref. 1) -- 3). 
This enables us to see how the ``S-matrix'' 
extracted from the two-point function 
can be diagonalized by Schur polynomials.

We focus on 
$Z=\frac{1}{\sqrt{2}}(\phi_1+i\phi_2)$, where 
$\phi_{1}$ and $\phi_{2}$ are two of the six adjoint scalar fields 
in $\Ncal$=4 SYM. As known from the non-renormalization theorem, 
the extremal correlator 
\begin{align}
\big\langle  
\prod_{i=1}^S\Tr {Z^\dagger}^{J_i}(y) \prod_{j=1}^P\Tr Z^{J_j}(x_1)\cdots 
\prod_{k=1}^Q\Tr Z^{J_k}(x_r)\big\rangle 
\label{co}
\end{align}
has no dependence on the gauge coupling. 
When the theory is compactified on $R\times S^3$, 
the lowest KK modes 
depend only on time, and it is possible 
to use 
the complex matrix model, whose action is given by 
\begin{align}
\int dt\Tr \big[\dot Z^\dagger (t) \dot Z(t) - Z^\dagger (t) Z(t)\big],
\label{La}
\end{align}
to evaluate Eq. \eqref{co} in the free field limit. 
Let us also explain why 
the Hamiltonian $H$ corresponding to Eq. \eqref{La} can be diagonalized 
using Schur polynomials. Due to 
the VanderMonde determinant arising from 
the measure $dZdZ^\dagger$, 
we can absorb it into the 
wave function $\Psi$ to redefine 
\begin{align}
\Psi ~\rightarrow~\triangle \Psi,
&&H~\rightarrow~\triangle H \frac{1}{\triangle},
&&\triangle=\prod_{1\le i<j\le N}(\lambda_i-\lambda_j).
\end{align}
The eigenstate of $H$ is of the form 
\begin{align}
\Psi=\chi_0 \det_{i,j} \lambda_j^{n_i},
~~~~~\chi_0=e^{-\sum_i \lambda^\ast_i \lambda_i},
~~~~~i,j=1,\cdots ,N,
\label{Psi}
\end{align}
which is just the Slater determinant of $N$ fermions in 
a harmonic oscillator potential. 
Note that the ground state is $\Psi_0=\chi_0 \triangle$, and the 
inner product now becomes 
$\langle\Psi|\Psi \rangle=\int d\lambda^\ast d\lambda ~\Psi^\ast
\Psi$ without the factor $|\triangle|^2$. 
A tool of particular use is the Weyl character formula
\begin{align}
S(\vec{r})=\frac{\det_{1\le i,j\le N} \lambda_j^{N-i+r_i}}{\triangle},
\label{Sch}
\end{align}
where $S(\vec{r})$ is the Schur polynomial. 
Here, $\vec{r}=(r_1,\cdots,r_N)$ represents the row lengths of a Young diagram $R$, 
which assigns a representation of U($N$) 
or the symmetric group $\text{S}_{n}$ ($n=\sum r_i$). 
Setting $n_i=N-i+r_i$ and 
using Eq. \eqref{Sch}, we can rewrite an excited state $\Psi$ as a product of the 
ground state and a Schur polynomial, i.e. 
$\Psi=\Psi_0 S(\vec{r})$. The 
energy eigenvalue of this excited state is 
\begin{align}
\sum^N_{i=1}n_i=n+\frac{N(N+1)}{2}, 
\label{energy}
\end{align}
where $n$ stands for 
the U(1) R-charge.

Let us return to the two-point function 
\begin{align}
\big\langle  \prod_{k=1}^P\Tr {Z^\dagger}^{J_k}(t)
 \prod_{l=1}^Q\Tr Z^{J_l}(t')\big\rangle 
 ={G}e^{iJ(t' - t)},
~~~~~~~J=\sum_{l} J_l=\sum_{k} J_k,
\label{two}
\end{align}
where 
$G$ (the ``S-matrix'' of scattering $\{J_k \}\to\{J_l \}$) is expressed 
as\cite{oku,Tsu,je1} 
\begin{align}
\begin{aligned}
G(\{J_k \};\{J_l \})=\int dZ d{Z^\dagger}~ e^{-2\text {Tr}(Z^\dagger Z)}
\prod_{k=1}^P\Tr {Z^\dagger}^{J_k}\prod_{l=1}^Q\Tr Z^{J_l}.
\label{G}
\end{aligned}
\end{align}
From Eq. \eqref{Sch}, it is found that 
$G$ is diagonalized as%
\footnote{We have omitted an overall factor resulting from off-diagonal 
elements of $Z$.}
\begin{align}
\begin{aligned}
G^{diag}=
&\int d\lambda^\ast d\lambda ~\Psi_{\{n_i\}}^\ast
\Psi_{\{n'_i\}}\\
=&\int dZ d{Z^\dagger}~ e^{-2\text {Tr}(Z^\dagger Z)}
S_S (Z^\dagger) S_R (Z)=t(R)\delta_{SR},
\label{Or}
\end{aligned}
\end{align}
where $S_R (Z)=\langle R|Z \rangle$ denotes the Schur polynomial. 
Moreover, $t(R)$ in Eq. \eqref{Or} has been 
determined by Jevicki et al. in Ref. 3) to be 
\begin{align}
t(R)=\frac{dimR(N)}{d(R)}=\prod_{\square (i,j)}(N-i+j), ~~~~~~d(R)=
\prod_{\square (i,j)}\frac{1}{h_{i,j}},
\end{align}
where $\square(i,j)$ and $h_{i,j}$ label the location and 
the hook length 
of the box $\square$, respectively, in the Young diagram.

For later convenience, 
we show that the above $|R\rangle$ can be written in terms of 
relativistic fermions. 
As in Ref. 3), we can rewrite the Schur polynomial as
\begin{align}
S_R (Z)=\sum_{\vec{k}} \langle R|\vec{k}\rangle \langle \vec{k}|Z \rangle
=\frac{1}{n!}\sum_{\sigma\in\text{S}_n}\chi_R (\sigma)\Tr(\sigma Z),
\label{Sr}
\end{align}
where the total box number $n$ 
is related to 
$\vec{k}=\{k_\ell\}$ 
by 
$n=\sum_\ell \ell k_\ell$, while 
$\chi_R(\sigma)$ denotes the character of the permutation $\sigma$. 
Then, introducing the 
coherent state representation 
\begin{align}
|Z\rangle =e^{\sum_{n>0}\frac{1}{n}(\Tr Z^n) \alpha_{-n}}|0\rangle,
\end{align}
we have 
\begin{align}
\langle \vec{k}|Z \rangle\equiv\prod_{\ell}(\Tr Z^\ell)^{k_\ell},
&&|\vec{k}\rangle=\prod_{\ell>0}(\alpha_{-\ell})^{k_{\ell}}|0\rangle,
&&[\alpha_m,\alpha_n]=m\delta_{m+n,0}.
\end{align}
By further applying the 
fermionization, i.e. 
\begin{align}
\alpha_n=\sum_{r\in{\mathbb Z}+1/2}b_r c_{n-r}, ~~~~~~
~\{c_r,b_s\}=\delta_{r+s,0},
\label{bcc}
\end{align}
the linear combination 
$\sum_{\vec{k}} \langle R|\vec{k}\rangle \langle \vec{k}|$ 
gives 
\begin{align}
|R \rangle
=\prod^{diag(R)}_{i=1}c_{-r_i +i-\frac{1}{2}} ~
b_{-h_i +i-\frac{1}{2}}|0\rangle, &&
b_s|0\rangle =c_s|0\rangle =0, ~~~~~s>0.
\label{fb}
\end{align}
Here, $r_i$ $(h_i)$ is the $i$-th row (column) length of 
the Young diagram $R$, while 
$diag(R)$ is the 
diagonal box number. 
These fermions are identically those appearing 
in 2d YM on a cylinder if we identify $c_{-r_i +i-\frac{1}{2}}$ 
with the $i$-th 
particle at level $(r_i -i +1)$ above 
the Fermi level $n_F$, and 
$b_{-h_i +i-\frac{1}{2}}$ 
with the $i$-th hole 
at level $(h_i-i)$ below $n_F$. 
Due to \eqref{bcc}, 
discrete R-charges (quantum numbers of $\alpha$'s) can thus be 
mapped to momenta of 2d relativistic free fermions.

The authors of Ref. 5) 
summarized the leading planar result for $G$ given in \eqref{G} with a 
graphical method. For example, the $1\to 4$ case is shown to be 
\begin{align}
\begin{aligned}
G(J;J_1,J_2,J_3,J_4)=
J_1 J_2 J_3 J_4 J(J-1)(J-2)N^{-3}.
\label{GJ}
\end{aligned}
\end{align}
We will see below that Eq. \eqref{GJ} can be reproduced by 
the tree-level tachyon scattering in the c=1 matrix model.

\section{c=1 matrix model and scattering of tachyons}
Let us briefly review some basics of the c=1 matrix model. 
The Lagrangian is defined as 
\begin{align}
\int dt\Tr\big
[\frac{1}{2}(D_t \Phi)^2 + \frac{1}{2}\Phi^2  \big], ~~~~~~
D_t \Phi=\partial_t \Phi+[A_t,\Phi],
\label{Sin}
\end{align}
where $\Phi$ is an $N\times N$ Hermitian matrix and 
the non-dynamical gauge field $A_t$ is introduced in order to 
project the wave function $\Psi(\Phi)$ onto the singlet sector. 
It is well known that Eq. \eqref{Sin} is equivalent to a fermion liquid in an upside-down harmonic
oscillator potential. 
The Hamiltonian is given by 
\begin{align}
H=\frac{1}{2\pi}\int dx\int_{p_{-}}^{p_{+}} dp ~\frac{1}{2}(p^2 - x^2),
\end{align}
where we have 
$p_{\pm}(x,t)=\pm\sqrt{x^2-2\mu}$, and 
$-\mu$ denotes the Fermi level below the tip of the potential. 
Defining $x=-e^{-q}$ (i.e. $x\in[-\infty,0]$ and $q\in[-\infty,\infty]$), 
we can further set 
\begin{align}
\begin{aligned}
p_{\pm}(q,t)=\pm e^{-q} \mp\epsilon_{\pm}(q,t)e^{q}, ~~~~
\epsilon_{\pm}=\sqrt{\pi}(\pm\Pi_S -\partial_{q} S),
\end{aligned}
\end{align}
so that 
\begin{align}
H= \frac{1}{2}\int dq ~\big[\Pi^2_{S} + (\partial_{q} S)^2 + 
e^{2q}{\Ocal}(S^3)\big].
\end{align}
That is, $S$ describes fluctuations (ripples) on the 
Fermi surface.

For $q\to-\infty$, $S(q,t)$ behaves asymptotically 
like a massless field and can be 
expanded as 
\begin{align}
S(q,t)=\int^{\infty}_{-\infty}\frac{d\xi}{2\sqrt{\pi} |\xi|}
\left(a_\xi e^{-i|\xi|t+i\xi q} +a^\dagger_\xi e^{i|\xi|t-i\xi q}\right),
&&[a_\xi,a^\dagger_{\xi'}] = |\xi|\delta(\xi-\xi').
\label{SS}
\end{align}
Then, by defining that 
\begin{align}
|\xi; in \rangle = a^\dagger_{\xi} |0\rangle, ~~~~\xi >0,
\end{align}
the tachyon scattering amplitude can be computed as follows.%
$\footnote{Here, the name ``tachyon'' 
arises from the fact that usually $e^{2q}S(q,t)$ is regarded as a tachyon field 
$T(q,t)$ (up to a phase) in the dual Liouville theory, see Ref. 6).}$ 
According to Polchinski\cite{Pol}, the tree-level 
tachyon scattering 
can be calculated by means of the hidden ${\cal W}_{\infty}$ symmetry. 
Due to the Liouville wall, incoming and outgoing modes are related as 
\begin{align}
a^\dagger_\xi = 
(\frac{1}{2}\mu)^{-i\xi} \sum_{n=1}^\infty 
\frac{1}{n!} \left(\frac{i}{\mu}\right)^{n-1} 
\frac{\Gamma(1-i\xi)}{\Gamma(2-n-i\xi)} 
\int_{-\infty}^0 d^n\xi_\ell 
\prod^n_{\ell=1}(a_{\xi_\ell}^\dagger -a_{\xi_\ell})~
\delta(\sum^n_{\ell=1} \pm |\xi_\ell| -\xi),
\label{are}
\end{align}
where the sign of $|\xi_\ell|$'s is plus (minus) 
if the creation (annihilation) operator is chosen 
in front of the delta function. 
The $1\to n$ amplitude is (up to leg factors)
\begin{align}
\begin{aligned}
\langle \xi_1 \cdots \xi_n;out| \xi;in\rangle &
=  \left(\frac{i}{\mu}\right)^{n-1} 
\frac{\Gamma(1+i\xi)}{\Gamma(2-n+i\xi)} 2\pi
\delta (\xi_1+\cdots +\xi_n -\xi)\prod_{\ell=1}^n \xi_\ell,\\
\frac{\Gamma(1+i\xi)}{\Gamma(2-n+i\xi)}&=(i\xi)(i\xi-1)\cdots(i\xi-n+2). 
\label{scat}
\end{aligned}
\end{align}
It is also well known that 
the S-matrix of the c=1 matrix model can be diagonalized using 
fermionic fields $b(z)$ and $c(z)$ (where $z$ is the complex coordinate) 
\begin{align}
b(z)=\sum_{r\in{\mathbb Z}+1/2}b_r z^{-r-1/2},&&
c(z)=\sum_{r\in{\mathbb Z}+1/2}c_r z^{-r-1/2},&&
&&\{c_r,b_s\}=\delta_{r+s,0},
\end{align}
which are related to the above non-relativistic fermions by the second 
quantization\cite{GK}. 
The reason is that the incoming mode $b_{-r}$ 
differs from the outing mode $({\cal R}b)_{-r}$ 
by a reflection factor ${\cal R}$\cite{Moore}, i.e. 
\begin{align}
{\cal R}_r=
 i\sqrt{\frac{1+ie^{-\pi(\mu +ir)}}
 {1-ie^{-\pi(\mu +ir)}}} \sqrt{\frac{\Gamma(\frac{1}{2}-i\mu+r)}
 {\Gamma(\frac{1}{2}+i\mu-r)}}. 
\end{align}
In other words, the c=1 matrix model is free in terms of the $b,c$ system, 
which is the fermionization version of the asymptotical $S(q,t)$ via 
\begin{align}
c\sim :e^{i\int^q dq'(\Pi_S -\partial_{q'} S)}:,~~~~~
b\sim :e^{-i\int^q dq'(\Pi_S -\partial_{q'} S)}:.~~~~~
\end{align}
Just as done in \eqref{bcc}, $\xi$'s of tachyons can be thus mapped 
to momenta of these 2d relativistic fermions.

Following Refs. 10) and 11), we see that 
$|R\rangle$'s in Eq. \eqref{fb} form a diagonal basis of the c=1 S-matrix 
such that 
\begin{align}
\begin{aligned}
\langle R|S_{c=1}|R \rangle
&=\prod^{diag(R)}_{i=1}\frac{\Gamma (i\mu+r_i-i+1)}{\Gamma(i\mu-h_i+i)}
\frac{\cos[\frac{\pi}{2}(r_i-i+i\mu)]\cos[\frac{\pi}{2}(h_i-i-i\mu)]}
{\sin[\pi(i\mu-h_i+i)]}+{\Ocal}(e^{-\mu})\\
&\approx e^{-\frac{i\pi}{2}\sum^{diag(R)}_{i=1}(r_i+h_i-2i)}~\frac{dimR(\mu)}{d(R)}. 
\label{Sc}
\end{aligned}
\end{align}
For large $\mu$, up to a pure phase, 
the diagonal element in the last line is just the 
aforementioned $t(R)$, if we 
replace $\mu$ with $N$.

Based on the two 
diagonalized S-matrix elements in Eqs. \eqref{Or} and \eqref{Sc}, 
we observe that through the prescription, i.e. $i\xi\to J$ and 
$\mu \to N$, Eqs. \eqref{GJ} and \eqref{scat} are identical, 
up to a delta function and an irrelevant phase. 
Therefore, 
we arrive at the same conclusion as the authors of Ref. 5), in which 
the Euclideanized AdS droplet approach is used to show this equivalence. 
The above observation can be understood as follows. 
For chiral primary operators (i.e. their conformal 
dimensions are equal to the R-charges), 
which can be treated as $N$-fermion states in a 
harmonic oscillator potential, 
in the phase space 
we are thus equivalently comparing the following two kinds of fermion liquids: 
\begin{align}
\begin{aligned}
x^2-p^2&\ge2\mu,\\
x^2+p^2&\le 2N.
\label{ppp}
\end{aligned}
\end{align}
The Fermi surface in the first case 
is determined by 
a hyperbola, 
while that in the second one is a circle of radius $\sqrt{2N}$. 
In addition, we recall that 
both $\xi$ (in the c=1 matrix model) and $J$ 
(R-charge in SYM) are related to momenta of 2d 
relativistic fermions, 
so 
$i\xi\to J$ 
can be interpreted as a result from 
the sign change of the ``non-relativistic'' $p^2$ term in Eq. \eqref{ppp}.

\subsection*{Acknowledgements}

We are grateful to Yutaka Matsuo, Yu Nakayama and Satoshi Yamaguchi 
for valuable comments. 

Note added: Near the completion of this work, we received 
a new preprint 
hep-th/0612262 from Jevicki and Yoneya, 
who reach the same conclusion through a detailed analysis of the 
Euclideanized AdS droplet scattering. 

\end{document}